\documentclass[%
 reprint,
superscriptaddress,
 amsmath,amssymb,
 aps,
]{revtex4-1}
\usepackage[normalem]{ulem}
\usepackage{color}
\usepackage{comment}
\usepackage{graphicx}
\usepackage{dcolumn}
\usepackage{bm}
\usepackage{graphicx}

\begin{document}
\newcommand{\lrl}[1]{{\color{green} [Lee: #1]}}
\newcommand{\yc}[1]{{\color{blue} [YC: #1]}}
\newcommand{\dr}[1]{{\color{red} [DR: #1]}}
\preprint{APS/123-QED}

\title{Observation of full contrast icosahedral Bose-Einstein statistics \\in laser desorbed, buffer gas cooled C$_{60}$}
\author{Ya-Chu Chan}
\affiliation{JILA, National Institute of Standards and Technology and University of Colorado, Boulder, Boulder, CO 80309}
\affiliation{Department of Chemistry, University of Colorado, Boulder, CO 80309}

\author{Lee R. Liu}
\email{lee.richard.liu@gmail.com}
\affiliation{JILA, National Institute of Standards and Technology and University of Colorado, Boulder, Boulder, CO 80309}
\affiliation{Department of Physics, University of Colorado, Boulder, CO 80309}

\author{Andrew Scheck}
\affiliation{JILA, National Institute of Standards and Technology and University of Colorado, Boulder, Boulder, CO 80309}
\affiliation{Department of Physics, University of Colorado, Boulder, CO 80309}

\author{David J. Nesbitt}
\affiliation{JILA, National Institute of Standards and Technology and University of Colorado, Boulder, Boulder, CO 80309}
\affiliation{Department of Chemistry, University of Colorado, Boulder, CO 80309}
\affiliation{Department of Physics, University of Colorado, Boulder, CO 80309}

\author{Jun Ye}
\email{ye@jila.colorado.edu}
\affiliation{JILA, National Institute of Standards and Technology and University of Colorado, Boulder, Boulder, CO 80309}
\affiliation{Department of Physics, University of Colorado, Boulder, CO 80309}

\author{Dina Rosenberg}
\affiliation{JILA, National Institute of Standards and Technology and University of Colorado, Boulder, Boulder, CO 80309}
\affiliation{Department of Physics, University of Colorado, Boulder, CO 80309}
\date{\today}

\begin{abstract}
  
The quantum mechanical nature of spherical top molecules is particularly evident at low angular momentum quantum number $J$. 
Using infrared spectroscopy on the 8.4~$\mu$m rovibrational band of buffer gas cooled $^{12}$C$_{60}$, we observe the hitherto unseen R($J=0-29$) rotational progression, including the complete disappearance of certain transitions due to the molecule's perfect icosahedral symmetry and identical bosonic nuclei. The observation of extremely weak C$_{60}$ absorption is facilitated by a
laser desorption C$_{60}$ vapor source, which transfers 1000-fold less heat to the cryogenic buffer gas cell than a traditional oven source. 
This technique paves the way to cooling C$_{60}$ and other large 
gas phase molecules to much lower temperatures, providing continued advances for spectral resolution and sensitivity.
\end{abstract}

\maketitle

Due to their strong intramolecular interactions, perfect rotational symmetries, and long-lived rotational and vibrational degrees of freedom, large molecules are interesting platforms for probing strongly correlated many body phenomena~\cite{Liu2023,Schulz2015} and quantum information processing~\cite{Sawant2020,zeppenfeld2023robust,albert2024topology}. 
Quantum state resolved infrared spectroscopy is the key to unlock these capabilities. Molecules beyond $\sim10$ atoms, however, typically exhibit intrinsic spectral congestion in the usual infrared bands due to the rapidly increasing vibrational density of states with increasing energy and atom number~\cite{Spaun2016,Changala2016}. C$_{60}$ is a notable exception: due to its rigidity and symmetry, it is the largest molecule for which quantum state resolution in the infrared was achieved \cite{Changala2019}. Based on this result, a series of high resolution spectroscopic measurements have provided the clearest understanding of structure and dynamics of individual C$_{60}$ molecules to date. By fitting high resolution spectra to effective molecular Hamiltonians, the spectroscopic constants for the ground and excited (T$_{1u}$(3)) vibrational states were determined \cite{Changala2019}. Then, detailed studies of the collisions between C$_{60}$ and a variety of atomic and molecular species explored the intramolecular pathways of energy relaxation through rotational and vibrational degrees of freedom \cite{Liu2022}. The observation of ergodicity breaking in the rotational fine structure of C$_{60}$ \cite{Liu2023} offers a new perspective on emergent phenomena in mesoscopic quantum systems made possible by free rotations, rotation-vibration coupling and perfect rotational symmetry. 

These results 
were obtained using a 900~K oven source coupled to a cryogenic buffer gas cell~\cite{Weinstein1998,Spaun2016,Patterson2007}. However, the significant heat load of the oven source constrained the rotational temperature to lie $\sim $150 K and above, preventing the concentration of significant population into states of low angular momentum quantum number $J$~\cite{Changala2019}. At low $J$, some rotational states are completely forbidden for the totally symmetric nuclear spin species. 
Because of the high symmetry and 60 identical spinless bosonic nuclei of $^{12}$C$_{60}$, these forbidden states are completely absent from its infrared spectrum, providing arguably the most striking example of symmetry and quantum mechanical effects on spherical top spectra~\cite{Bunker1999}. 

In this work, we develop a laser desorption vapor source to 
produce a colder sample of gas phase C$_{60}$. This allows us to measure the R branch rotational progression of $^{12}$C$_{60}$ from $J=0-29$ and beyond using cavity-enhanced high resolution spectroscopy on the 8.4~$\mu$m 
rovibrational band, demonstrating complete absence of rotational transitions, as predicted by nuclear spin statistics. 
We characterize the rotational temperature of laser desorbed C$_{60}$ and find that it reaches below 100~K using both Ar and He buffer gas cooling. 
Rotational and Doppler thermometry indicate that rotational and translational degrees of freedom are essentially thermalized. We also observe large signal from cold C$_{60}$ with He buffer gas, indicating that in spite of the enormous mismatch in masses, small He-C$_{60}$ rotational quenching cross section~\cite{Liu2022}, and highly non thermal optical excitation by 532~nm photons, He is able to efficiently cool the vibrational, rotational, and translational degrees of freedom of laser desorbed C$_{60}$.  Our specific laser desorption technique possesses several unique features that enable high sensitivity molecular spectroscopy: 
1) three orders of magnitude reduction in heat load on the cryogenic buffer gas cell compared to a traditional oven source; 
2) stable pulse-to-pulse desorption yield, varying by less than 10\%, due to the uniformity of the deposited C$_{60}$ film thickness; 3) the use of unusually long duration (200~ms) and low peak power (2~W) desorption pulses to minimize fragmentation of C$_{60}$, in contrast to MW peak powers for traditional ns pulsed lasers; 
4) efficient collisional cooling by He; and 5) controllable, transient ($\sim$30~ms response time) C$_{60}$ yield, enabling rapid subtraction of probe light intensity fluctuations from absorption signal.
\begin{figure}
\centering
	\includegraphics[width = \columnwidth, trim={4.75cm 2.9cm 2.75cm 0.5cm},clip]{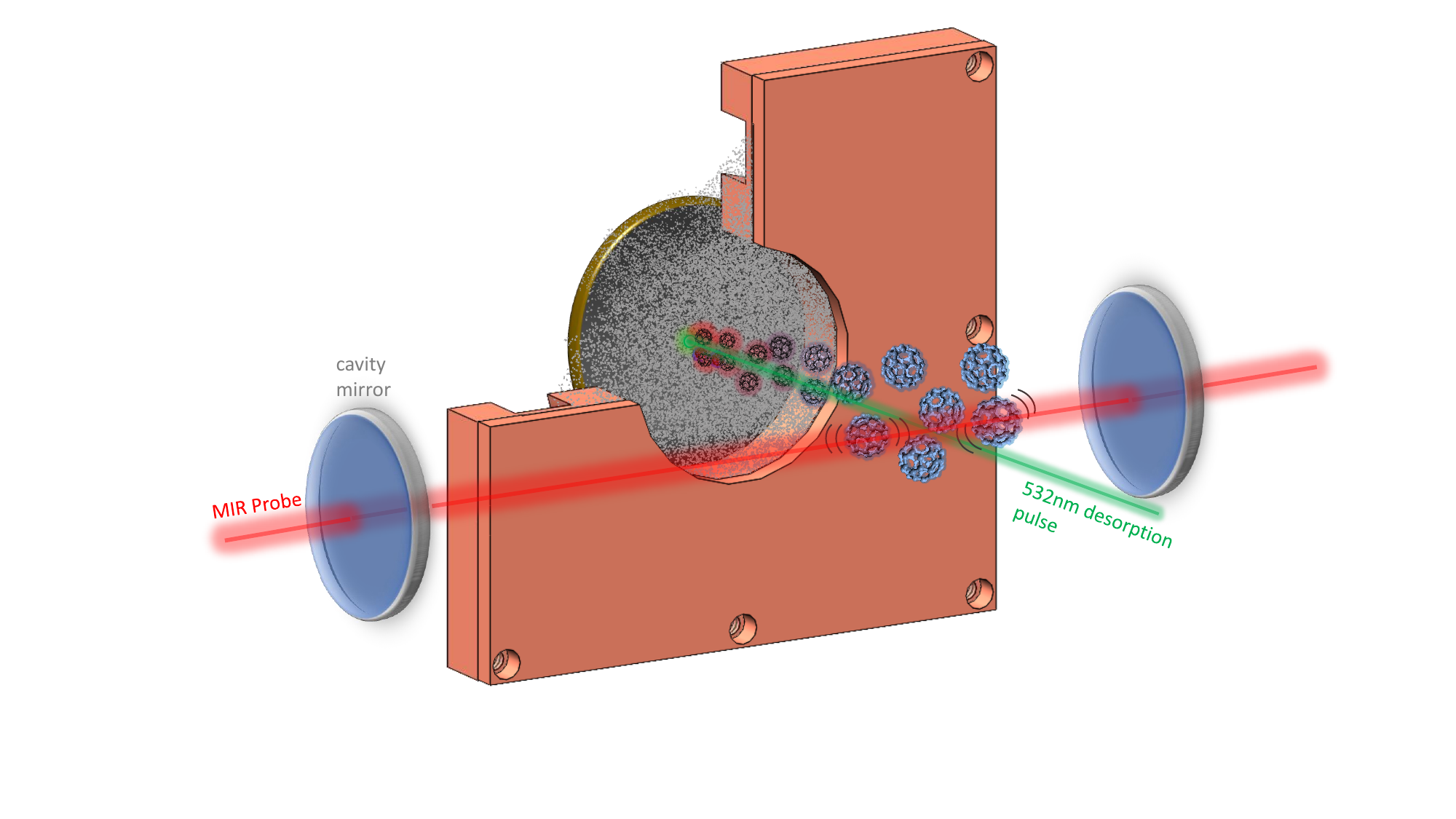}
	\caption{\textbf{Laser desorption and buffer gas cooling of C$_{60}$} 
 A 5~$\mu$m-thick $C_{60}$ film target is deposited on a tantalum foil via thermal evaporation. The target is held in the center of an annular buffer gas inlet. Buffer gas flow is directed over the face of the target.  A 532~nm cw desorption laser (green) is focused onto the target and rastered across in steps of 600~$\mu$m using a piezo-controlled mirror. At every point, the laser is pulsed on for 200~ms and desorbs a constant amount of C$_{60}$, determined by the thickness of the target and 300~$\mu$m focal spot size. Each 1-inch diameter target yields approximately 7,200 shots. The laser power is chosen to completely penetrate the C$_{60}$ film, thereby desorbing a nearly constant amount of C$_{60}$ into the cell. 
 The desorbed C$_{60}$ molecules cool as they diffuse through the buffer gas. The 8.4~$\mu$m mid infrared probe beam (red) resonantly excites a vibrational mode of C$_{60}$ and is absorbed. The mirrors form an optical cavity ($\mathcal{F}$ = 12,500) to enhance the absorption due to C$_{60}$. Not shown, surrounding the inlet is the buffer gas cell anchored to a liquid nitrogen cold finger, maintained at a temperature of 82~K.} 
	\label{fig:apparatus}
\end{figure}

The laser desorption setup is described in Fig.~\ref{fig:apparatus}. 
The heat load from laser desorption alone is bounded from above by the total time averaged input laser power of 1~W.  The laser desorption therefore represents a 1000-fold reduction in heat load compared to the traditional 1~kW, 900~K oven source~\cite{Changala2019}. 
Next, we characterize the shot-to-shot uniformity of the transient C$_{60}$ yield. Because of the low thermal conductivity of Ar, and because it was shown to efficiently cool C$_{60}$ in previous studies~\cite{Changala2019}, we primarily use Ar buffer gas cooling for characterization. 
We pulse the desorption laser at 2.5 Hz with 50\% duty cycle and record the transient absorption of the 8.4~$\mu$m probe laser parked at 1184.8475~cm$^{-1}$, where the C$_{60}$ absorption spectrum is relatively featureless due to spectral congestion of the Q branch. Remarkably, the peak absorption, which is proportional to the C$_{60}$ density, only fluctuates by $<$10\% from shot-to-shot. Because the desorption laser completely penetrates the C$_{60}$ film, the yield is largely dictated by the film thickness, which is uniform to within a few percent, consistent with the shot-to-shot variation in C$_{60}$ yield.

 \begin{figure}
\centering
	\includegraphics[width=\columnwidth]{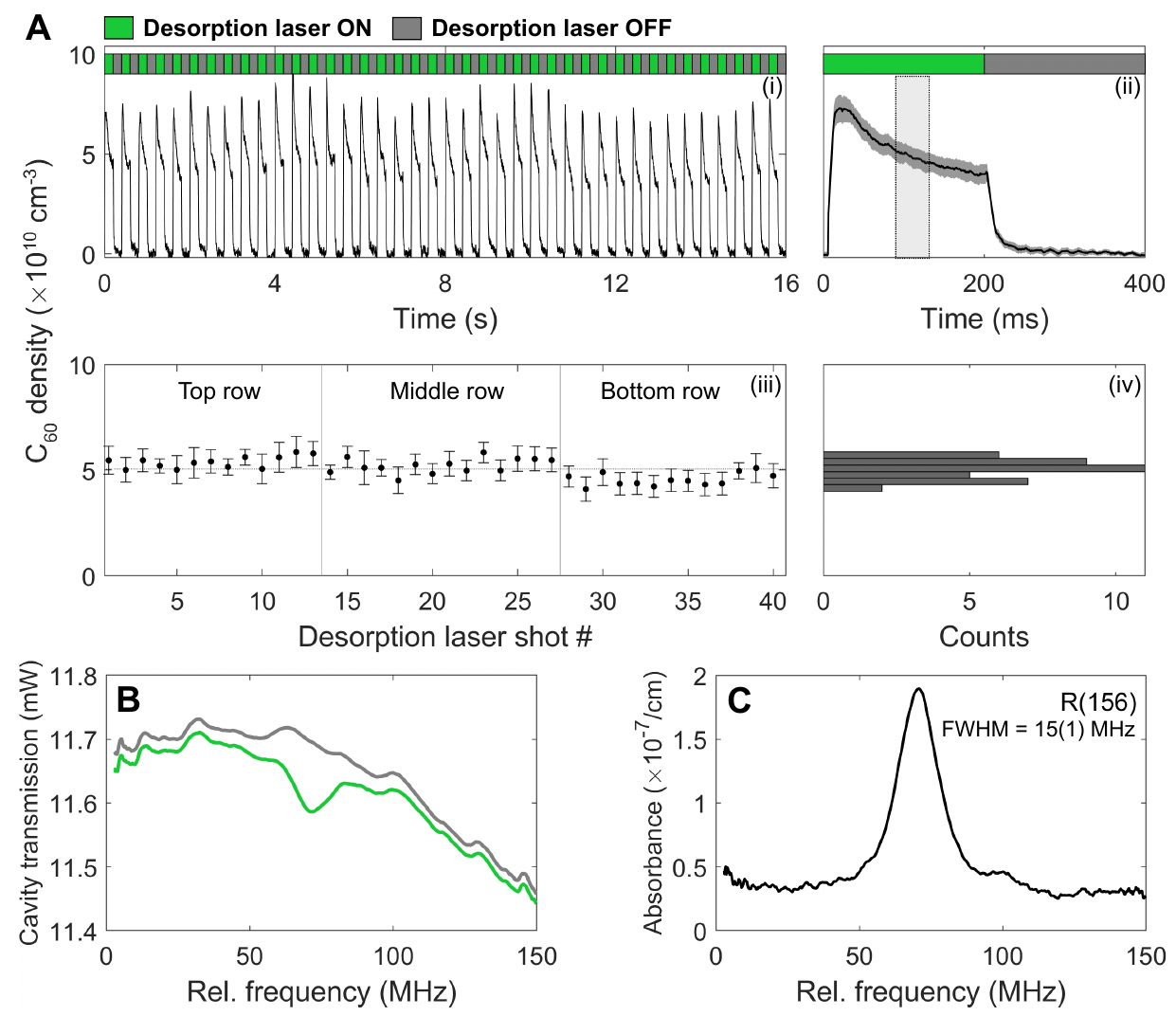}
	\caption{\textbf{Characterization of transient laser desorbed C$_{60}$ yields.} A)(i) Transient C$_{60}$ yield converted from absorption in the Q branch region. 
 The desorption laser pulsing sequence is depicted by alternating green and grey squares. (ii) C$_{60}$ density profile averaged over 40 shots. Gray error band indicates the point-wise standard deviation. Around 100~ms after the desorption laser fires, the variation in the density is less than 10\% over a period of 50~ms (vertical rectangle) and is stable enough for acquiring spectral features of cold C$_{60}$. (iii) Mean and standard deviation of C$_{60}$ density within the 50~ms window for each laser shot across three different sections of the target. (iv) Histogram of mean density, showing uniform C$_{60}$ yield. 
    B) The 8.4~$\mu$m probe laser frequency is scanned 150~MHz around R($J=156$) with the desorption laser on (green) and off (gray), yielding ``filled" and ``empty" cavity spectra, respectively. C) Background-free molecular absorption signal, obtained by subtraction of empty cavity transmission from filled cavity transmission. The data is averaged over 40 shots.   
    }
	\label{fig:characterization}
\end{figure}
Fig.~\ref{fig:characterization}A(ii) shows the time trace of C$_{60}$ density averaged over 40 desorption laser shots. After the desorption laser has been on for 5 ms, C$_{60}$ density 
increases and reaches its maximum within 30~ms, then decays over the course of several hundred milliseconds. The variation in the density is less than 10\% in the 50 ms window around 100 ms after the desorption laser fires. At this point, we scan the laser over 150 MHz to acquire absorption spectra in the R branch region. Over a scan spanning $\sim$15 MHz, corresponding to the spectral line width, the C$_{60}$ density changes by less than 1\% and the effects on the line shape can be neglected. The desorption laser is shut off and the C$_{60}$ density drops to zero within 30~ms, after which another ``empty cavity" spectrum is acquired (Fig.~\ref{fig:characterization}B). The rapid response time of C$_{60}$ density allows for subtraction of 
slowly fluctuating intensity drifts such as those due to drifting parasitic etalons. In fact, laser desorption itself is instantaneous, but the measured 10's of milliseconds response time is dominated by transport through the buffer gas and our particular experimental geometry. 
The resulting background-free molecular absorption spectrum for R($J=156$) is shown in Fig.~\ref{fig:characterization}C. From the measured absorbance, we calculate the cold $^{12}$C$_{60}$ density to be approximately $5(1)\times 10^{10}$cm$^{-3}$, based on the integrated intensity of the 8.4~$\mu$m band~\cite{Iglesias-Groth2011a}, rotational temperature of 100~K, and effective optical path length of $\sim$475~m.


Having characterized the yield of laser desorbed C$_{60}$, we now turn to C$_{60}$ rotational and Doppler thermometry. In particular, the 532~nm photons may deposit energy primarily into certain (e.g., electronic) degrees of freedom, initializing the C$_{60}$ molecules in a highly non thermal state~\cite{Burgess1988}. In contrast, an oven source produces molecules in which all internal degrees of freedom are approximately thermalized. Therefore, a significant concern was that laser desorbed C$_{60}$ might not be efficiently cooled. On the other hand, since vibration-to-translation energy transfer is expected to be more favourable for smaller vibrational energy gaps, larger molecules might be expected to vibrationally quench easily due to collisions~\cite{Du1991}. Furthermore, thermalization may occur rapidly in larger molecules even in the absence of collisions due to efficient intramolecular vibrational energy redistribution (IVR)~\cite{Parmenter1982,Smalley1982,Smalley1983,Nesbitt1996}. 
Our measurements indicate that thermalization proceeds efficiently for laser desorbed C$_{60}$. A second key question was whether C$_{60}$ could be efficiently quenched by collisions with He instead of Ar. Previous work with a 900~K oven source of C$_{60}$ found that Ar gave the largest cold C$_{60}$ peak absorbance~\cite{Changala2019} and that He has the smallest rotational quenching cross section amongst all noble gases~\cite{Liu2022}. Furthermore, He is $10\times$ more thermally conductive than Ar, exacerbating the heat transfer from the 1~kW oven to the nearby cryogenic buffer gas cell. Nevertheless, He is highly desirable as a buffer gas due to its significant vapor pressure at 4~K, making it compatible with a helium cryostat. He buffer gas cooling is also expected to mitigate possible cluster formation at lower temperatures~\cite{Klos2024}. Therefore, it was crucial to prove that He buffer gas efficiently collisionally quenches laser desorbed C$_{60}$ as well. 


The liquid nitrogen cold finger and buffer gas cell temperatures were directly measured with Cernox thin film resistance cryogenic temperature sensors and silicon diodes to be 82~K. Residual heat loads, such as radiative and conductive coupling to the room temperature dewar walls, dominate the total heat load and prevent the temperature from reaching 77~K, even with the desorption laser inactive. The static pressure in the cell was measured directly with a capacitance manometer. Absorption spectra were measured for six $J$-values scattered across the thermally populated states R($J=66, 96, 126, 156, 186, 206$) at various buffer gas pressures ($330-410$~mTorr), which we controlled by varying the buffer gas flow rate ($80-130$~sccm). 

\begin{figure}
\centering
	\includegraphics[width = \columnwidth]{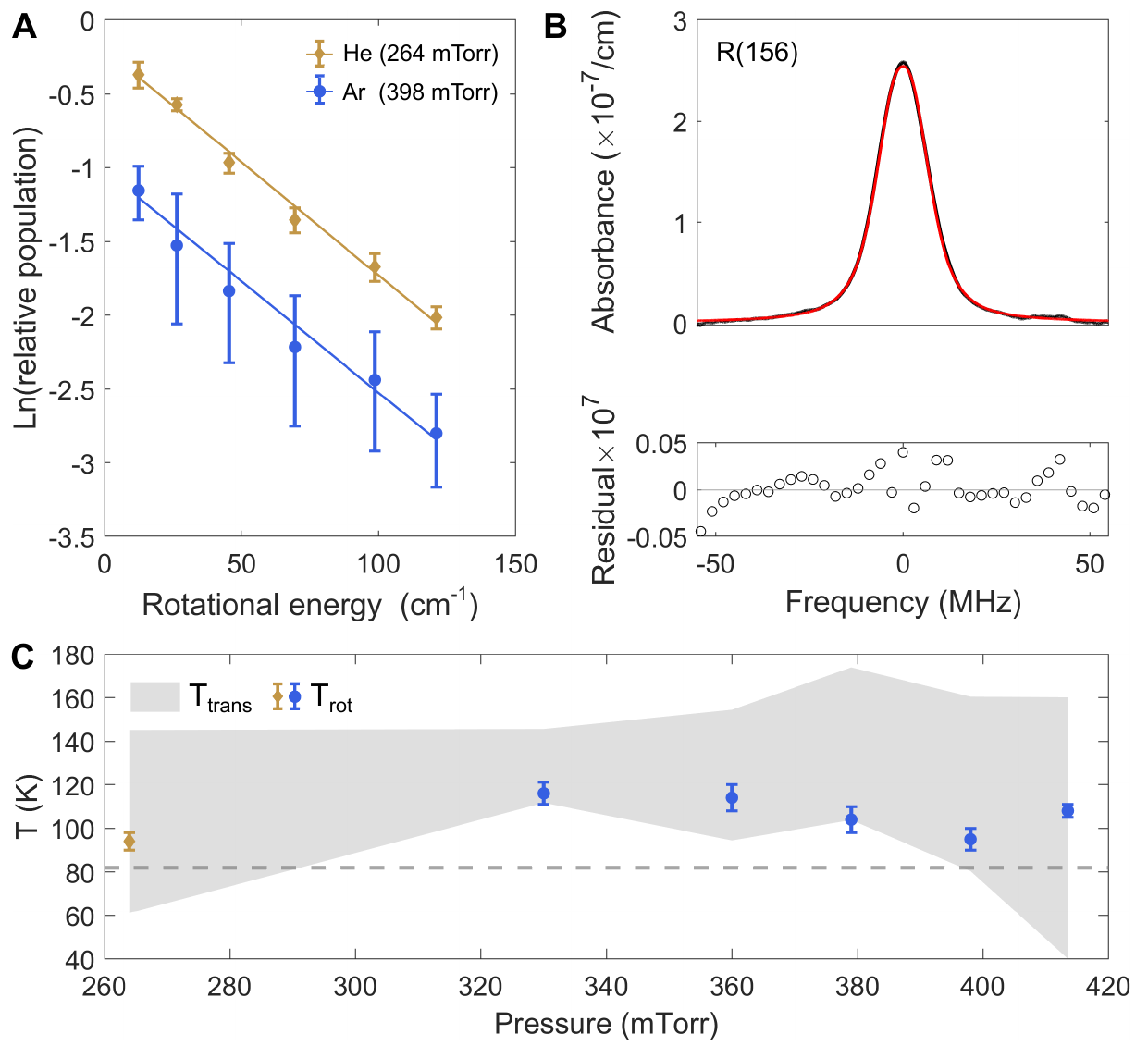}
	\caption{\textbf{Pressure-dependent Doppler and rotational thermometry of laser desorbed C$_{60}$ in He and Ar buffer gas.} A) Exemplary rotational temperature fits for C$_{60}$ in Ar and He buffer gas. 
 Note that He buffer gas cooling yields 
 higher and more uniform molecular signal (as evident from the smaller error bars in the integrated absorption).  B) Exemplary fit of R($J=156$) absorption to rate equation, with residuals plotted below.  C) Rotational and translational temperatures from Boltzmann analysis and rate equation fitting respectively, for a range of buffer gas pressures. The rotational temperatures lie within the error bounds of the fitted translational temperature. In addition, the coldest rotational temperatures approach  82~K, the temperature of the liquid nitrogen cold finger (horizontal dashed line). Note that He buffer gas cooling achieves the lowest rotational temperature, even at much lower pressure. }
	\label{fig:thermometry}
\end{figure}

The rotational temperature was obtained from the integrated absorbance at each of the six measured R($J$) lines. Assuming negligible population in the vibrationally excited states, the integrated absorbances $I(J)$ are dominated by the rotational populations in the vibrational ground state. Hence, $I(J) =C S_J g_{J} e^{-E_{J}/k T_{rot}}$, where $C$ is proportional to the total C$_{60}$ population, $S_J$ is the intrinsic line strength factor~\cite{Herzberg1945}, $g_{J}$ is the angular momentum degeneracy factor including lab-frame and molecular frame degeneracies for a spherical top and nuclear spin statistics weight~\cite{Bunker1999}, $E_J = BJ(J+1)$ is the rotational energy in the vibrational ground state, $k$ is the Boltzmann constant, and $T_{rot}$ is the rotational temperature. Rearranging this equation gives 
$\textrm{Ln}(I(J)/ (S_J g_{J})) = \textrm{Ln}(C) -E_{J}/(k T_{rot})$. In this way $T_{rot}$ can be extracted from the slope of a best-fit line to the logarithm of $I(J)/ (S_J g_{J})$, the relative populations per $J$ multiplet component (Fig.~\ref{fig:thermometry}A). For Ar buffer gas, we find that $T_{rot}$ decreases with increasing pressure up until 400~mTorr, after which it rises again (Fig.~\ref{fig:thermometry}C). The initial decrease in $T_{rot}$ is  expected: as pressure increases, the number of collisions experienced by C$_{60}$ 
increases, thereby lowering the measured $T_{rot}$.

The rise in T$_{rot}$ above 400~mTorr is also well understood. The total number of collisions a C$_{60}$ molecule experiences before reaching the 8.4~$\mu$m beam is proportional to the product of buffer gas pressure and the time of flight from the target. Separately, we observed a 
decrease in the time of flight with increasing buffer gas flow, suggesting that buffer gas ``carries" desorbed C$_{60}$ to the cavity axis at a speed that depends on the flow rate. Indeed, the time of flight ranges from a few to 10's of milliseconds depending on the flow rate, which is intermediate between ballistic and diffusive transport. Therefore, while higher buffer gas flow increases pressure, it also decreases the C$_{60}$ time of flight. Pressure in the cell begins to saturate at higher flow rates, possibly due to hydrodynamic effects~\cite{Patterson2007,Barry2011}. At sufficiently high flow rates therefore, the reduction in collision events due to time of flight overcomes the increase in collision rate due to pressure, and the total collision number drops with increasing flow rate. This is responsible for the measured increase in $T_{rot}$ 
above 400~mTorr. 

Since a considerable amount of averaging was needed to obtain sufficiently high signal to noise ratio for accurate line shape fitting, we obtained the translational temperature $T_{trans}$ by globally fitting spectra measured for three selected $J$ values ($J=96, 156, 186$) to line shapes predicted by a semi-classical steady-state rate equation model discussed in Ref~\cite{Liu2022}. In particular, the model accounts for optical saturation of two counter propagating laser beams and $J$-dependent rotational quenching, which were crucial to obtain convergence of $T_{trans}$ beyond simply fitting each line shape to an independent Voigt profile. We fix C$_{60}$-Ar collision cross section values from Ref~\cite{Liu2022} and allow the buffer gas temperature T$_{BG}$ and $T_{trans}$ to float, while constraining them to be equal (T$_{BG}$=T$_{trans}$) 
and independent of $J$. Here, T$_{BG}$ affects the diffusion rate of C$_{60}$ and the collision frequency. At our buffer gas pressures, we expect translational thermalization of C$_{60}$ to occur in less than a hundred collisions~\cite{Friedrich2009}, or a few microseconds, much shorter than the 10~ms timescale it takes for laser desorbed C$_{60}$ to reach the interaction region. For each pressure value, $T_{rot}$ was fixed by the independent rotational temperature fit described earlier.  An exemplary fit for R($J=156$) is shown in Fig.~\ref{fig:thermometry}B. 

For Ar buffer gas, we find that T$_{trans}$ falls in the range $130\pm 30$~K, with flat dependence on buffer gas pressure. The error bars reflect the fact that line shape fitting is generally insensitive to translational temperature, since inhomogeneous linewidth scales approximately as $\sqrt{T_{trans}}$. The translational and rotational temperatures agree within error bars and approach that of the liquid nitrogen cold finger, suggesting that rotational and translational degrees of freedom are nearly thermalized with each other and the buffer gas. We repeated the thermometry with He at a pressure of 265~mTorr and find that the fitted translational $T_{trans} = 103(42)$~K and rotational temperatures $T_{rot} = 94(4)$~K rival the lowest temperatures achieved with Ar, in spite of the much lower He pressure used (Fig.~\ref{fig:thermometry}C). He buffer gas cooling also yielded a signal of cold C$_{60}$ comparable to the highest C$_{60}$ signal achieved with Ar cooling, as well as fractionally more stable C$_{60}$ yield, indicated by the smaller error bars in (Fig.~\ref{fig:thermometry}A). We attribute this stability to the suppression of C$_{60}$-buffer gas cluster formation. This hypothesis is supported by the shallow interaction potential and lower near-dissociation threshold density of states of the C$_{60}$-He complex compared to C$_{60}$-Ar~\cite{Klos2024}. The large signal achieved with He buffer gas cooling also indicates that He collisions efficiently quench C$_{60}$ \textit{vibrations}, since transitions from excited vibrational states (vibrational ``hot bands") are expected to be red-shifted from transitions from the ground vibrational state (fundamental band) by $\sim$60~cm$^{-1}$~\cite{Wang2018}, and will not contribute to the resolved R branch features. With a greatly reduced heat load, He buffer gas cooling is fully compatible and in fact preferable for laser desorbed C$_{60}$.

As mentioned earlier, the apparently efficient thermalization of laser desorbed C$_{60}$ should not come as a surprise. 
Intramolecular redistribution, particularly in a molecule as large as C$_{60}$, might redistribute the internal energy rapidly amongst internal degrees of freedom, even without buffer gas cooling. Indeed, thermal C$_{60}$ sublimated from a 900~K oven source possesses 7.4~eV of vibrational energy, corresponding to the energy of \textit{three} desorption photons at 532~nm. Smaller molecules, on the other hand, are expected to exhibit quenching bottlenecks due to large energy gaps and sparse spectrum of internal states~\cite{Sinha1990,Sinha1992}. 
Our results are promising for laser desorption and subsequent collisional thermalization of other large molecules~\cite{Spaun2016}.

 \begin{figure*}
\centering
	\includegraphics[width=\textwidth]{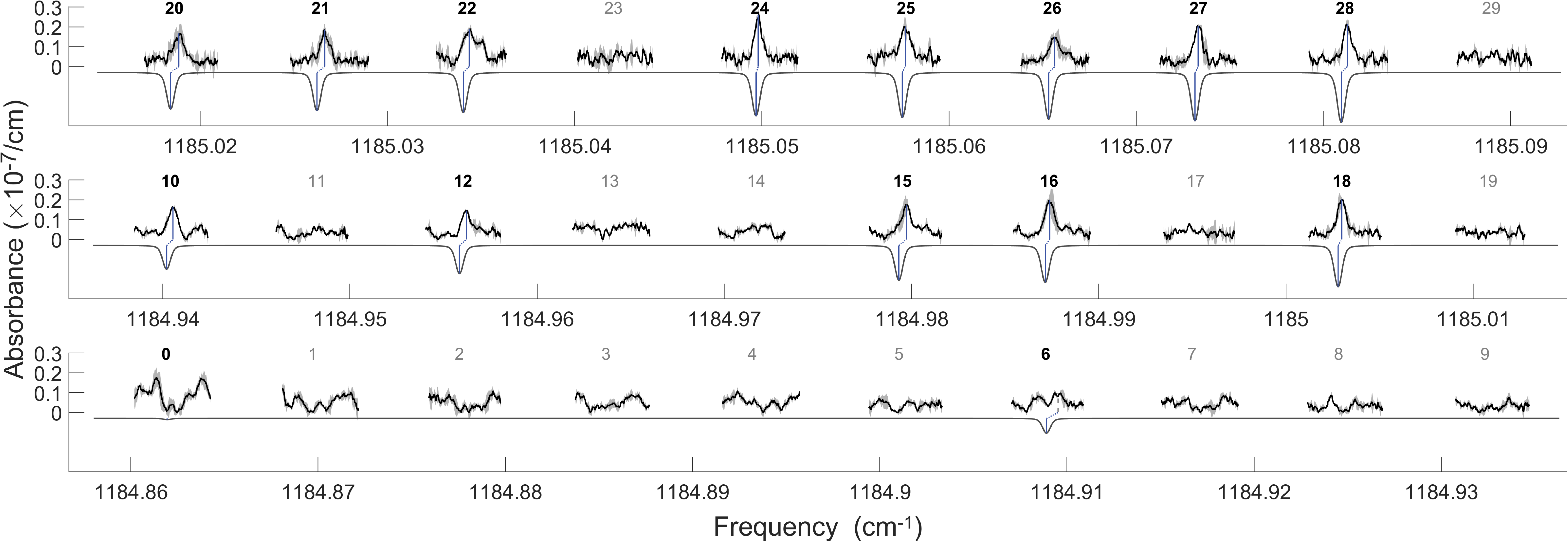}
	\caption{\textbf{Complete R($J=0-29$) absorption spectrum in the 8.4~$\mu$m rovibrational band of cold C$_{60}$.} 
 Experimental data (black) is the average of 200-400 raw spectra, with the light gray error bands representing the point-wise standard deviation. Simulated spectrum (gray, inverted) derived from spectroscopic constants determined in Ref~\cite{Changala2019}, assuming a rotational temperature of 100~K and including nuclear spin weights. Notably, the complete absence of many R($J$) transitions, as predicted by bosonic exchange statistics and the molecule's icosahedral symmetry, is clearly visible. There is an increasing $\sim5-15$~MHz blue-shift of experimental line centers (gray vertical lines) from the simulated spectrum as $J$ decreases from $29-0$, continuing the trend seen in Ref~\cite{Changala2019}. Note, we tentatively assign R($J = 6$) to the higher frequency peak in the doublets (marked with a dash vertical line). Additional features at $J=6$ and 22 and at very low $J$ 
 are deemed to be real. 
 } 
	\label{fig:missing_Js}
\end{figure*}

Having demonstrated efficient He buffer gas cooling of laser desorbed C$_{60}$, as well as the stable shot-to-shot C$_{60}$ yield, we now turn to observing the extremely weak molecular absorption in the hitherto unseen R($J=0-29$) spectral region. For $^{12}$C$_{60}$, since each carbon atom is a spinless, indistinguishable boson, the only allowed rotational states in the vibrational and electronic ground state are those which correlate to the totally symmetric irreducible representation in the icosahedral group. These correspond to $J = 0,6,10,12,\cdots$, with the remaining R($J$) transitions absent~\cite{Bunker1999}. 
The striking icosahedral nuclear spin weights arise from the two-, three-, and five-fold discrete rotational symmetry axes of $^{12}$C$_{60}$, and are borne out in our observed spectra (Fig.~\ref{fig:missing_Js}). 
However, we also observe perturbations to the transition intensities and frequencies, 
perhaps due to high-$J$ features from the Q branch, some accidentally degenerate vibrational combination state, or contamination from C$_{60}$ isotopologues. 
%
The observation of these subtle perturbations 
is facilitated by the He buffer gas cooling, which provides stable shot-to-shot C$_{60}$ yield and enables averaging of extremely weak molecular absorption signals.

In conclusion, we have demonstrated efficient cooling of laser desorbed C$_{60}$ to below 100~K, which is sufficiently cold to observe the hitherto unseen R($J=0-29$) transitions. This spectral region could be used for new tests of bosonic indistinguishability involving highly symmetric molecules~\cite{Angelis1996,Modugno1998,Mazzotti2001,Pastor2015}. In addition, the three orders of magnitude reduction in heat load of the laser desorption source compared to the oven source opens the door for cooling using a closed cycle He cryostat, which has much less cooling power than evaporating liquid nitrogen. At the 4~K temperature achievable by a He cryostat, C$_{60}$ population in the first 30 rotational levels in the electronic-vibrational ground state increases by two orders of magnitude, opening the door to understanding the origin of spectral perturbations, as well as providing a tantalizing route toward quantum state preparation of gas phase C$_{60}$ at high density. We note that large molecules such as C$_{60}$ are not amenable to the traditional approach of laser cooling due to their large mass (requiring correspondingly many more photon scattering events) and density of vibrational states (increasing the number of lasers required to ``repump" vibrational populations), although a scheme involving  chemical functionalization of a fullerene has been theoretically proposed~\cite{Klos2020}. 
Thus, laser desorption of C$_{60}$ opens up a new class of molecules to quantum science.
\\

We thank Adam Ellzey for technical assistance; the COSINC Fabrication facility at CU Boulder, particularly Dylan Bartusiak; and Jutta Toscano for the inspiration to pursue laser desorption of C$_{60}$. We thank Jacob Higgins and Baruch Margulis for providing useful comments on the manuscript. Funding support for this work is provided by AFOSR (FA9550-19-1-0148); NSF QLCI (QLCI OMA-2016244); NSF (Phys-1734006) and NIST. Support is also acknowledged from the U.S. Department of Energy, Office of Science, National Quantum Information Science Research Centers, Quantum Systems Accelerator. D.R. acknowledges support from the Israeli CHE Quantum Science fellowship, and is an awardee of the Weizmann Institute National Postdoctoral Award Program for Advancing Women in Science. Y.-C.C. and D.J.N. acknowledge support from the DOE (DE-FG02-09ER16021) and NSF (CHE 2053117).

\bibliography{desorption,yachu,arxiv}

\end{document}